# First Observed Metal-like to Insulator Transition in the vacant $3d$ orbital Quantum Spin Liquid $Tb_2Ti_2O_7$

B. Santhosh Kumar and C. Venkateswaran*

Department of Nuclear Physics, University of Madras, Guindy campus, Chennai – 600 025, India.

*\* E mail: cvunom@hotmail.com*

**Abstract**

We report the observation of metal-like to insulator transition (MTI) in the $3d$-pyrochlore oxide $Tb_2Ti_2O_7$ at 603 K due to the interaction of empty $3d$ orbitals of $Ti^{4+}$ with $O^{2-}$ ions, evidenced by the transition in resistivity. Magnetisation, specific heat capacity and differential scanning calorimetry support the MTI, and the transition is of second order. An appreciable change in magnetisation with temperature, without any magnetic phase transition, is a behaviour typical in this family of compounds which is seldom observed in empty $d$ orbital pyrochlores. The possible mechanism that supports the MTI in $Tb_2Ti_2O_7$ is discussed. Subsequently, a broad change in magnetisation from 696 K is also seen. Thermogravimetric analysis confirms the observed MTI (603 K) and the broad change in magnetisation (696 K) are not due to oxygen vacancy.

## I. Introduction

A little frustration in the spin of a material gives remarkable properties also making its physics more interesting. This is true in geometrically frustrated materials which have unusual phenomena like the random alignment of spins, interaction with neighbouring atoms, corner sharing tetrahedra and many more[1-8]. The term 'geometrical frustrated' refers, the spin of individual ions are well ordered but with the absence of individual interaction due to the competing magnetic behaviour of neighbour ions[9]. A large number of compounds in the frustrated family of materials crystallize in the stoichiometry $A_2B_2X_7$ (X = F, O), where A is a rare earth trivalent ion and B is a transition metal ion (A = Tb, Y, Dy, Ho, Gd, Er, Pr and B = Mo, Ti, Sn, Ir, Ru, Os, Mn) with eight-fold and six-fold oxygen co-ordination, respectively[7, 10]. This class of compounds form two interpenetrating lattice network of corner sharing tetrahedra of A and B which makes them thermal and chemically inert [7, 10].

The electrons in the 4d/5d pyrochlore compounds are localised compared to the 3d pyrochlores due to its greater hybridisation with $O^{2-}$ anions. The higher spatial degree of 4d/5d wave function over 3d leads to a decrease in the intra-atomic Coulomb repulsion and increases the wave function of the elements/ compounds. The degree of correlation, in electron correlated systems, is defined as the ratio of intra-atomic Coulomb repulsion (Hubbard U) to the bandwidth (W) [11-13]. The electron wave function will increase from 3d to 5d leading to a larger bandwidth that follows the order $W_{5d} > W_{4d} > W_{3d}$ and the Columbic repulsion will decrease accordingly as $U_{5d} < U_{4d} < U_{3d}$. Typically, U/W >> 1 for 3d oxides, for metals U/W << 1 and for 4d/5d systems U/W ~ 1. This property makes the physics of 3d pyrochlores more interesting.

In a short review of $A_2B_2O_7$ pyrochlores, $Cd_2Re_2O_7$ is the first superconductor at 1 K [14-16], evident from the heat capacity measurements revealing a clear second and first order phase transition with a change in its electrical resistivity at 200 K and 120 K [14]. Consequently, the heat capacity of $Pr_2Ru_2O_7$ shows a λ-type divergence at 162 K which is due to the antiferromagnetic ordering of $Ru^{4+}$ ion in the B-site[17]. Even in the last decade, $Hg_2Ru_2O_7$ was shown to exhibit a metal to insulator transition (MIT) at 108 K by Wilhelm Klein et al. [18] and Ayako Yamamoto [19]. It was observed, in general, for most of the oxide pyrochlores A-site ion is responsible for low temperature magnetic properties and B-site ion is responsible for electrical transition [20].

In the pyrochlore $Cd_2Os_2O_7$, semiconductor to metal transition is reported by A.W. Sleigh. et al. in 1973 [21]. Later, in 2000 D. Mandrus. et al. discussed the same observation as metal to insulator transition [16].

In an ideal metal to insulator transition (MIT) the resistivity of the sample normally varies from $10^{-3}$ Ω m to $10^{10}$ Ω m. But, $Hg_2Ru_2O_7$ showing MIT at 108 K has resistivity value of approximately 1.8 mΩ cm at room temperature and it enters into the insulator phase below 108 K with resistivity value around 7 mΩ cm. The change in the order of resistivity is approximately 3.3. Similarly, $Nd_2Ir_2O_7$ exhibiting MIT at 36 K shows a change in the order of metal to insulating phase by 10. In $A_2Ir_2O_7$ (A=Sm, Eu) the MIT is at 117 K and 120 K respectively, and the change in the order of resistivity is approximately $10^3$.

### a. Effect of A-site ionic radii on the electrical property of pyrochlore

The ionic radii of A site in pyrochlore lead to many interesting electrical properties. For example, the iridates, $A_2Ir_2O_7$, shows a wide range in electrical resistivity from metal to insulator. If A= $Pr^{3+}$, $Nd^{3+}$, $Sm^{3+}$ and $Eu^{3+}$ whose ionic radii decreases from 1.13 Å to 1.087 Å are metals [22-24]. If A= $Gd^{3+}$, $Tb^{3+}$, $Dy^{3+}$ and $Ho^{3+}$, ionic radii varying from 1.078 Å to 1.041 Å are insulators [24-25]. The ionic radii of A cation decide the bond length and bond angle of Ir-O. This kind of variation is because the larger $A^{3+}$ ($Pr^{3+}$, $Nd^{3+}$, $Sm^{3+}$ and $Eu^{3+}$) cation promotes the electron transfer via Ir-O-Ir hybridisation that makes the compound metal. Smaller cation ($Gd^{3+}$, $Tb^{3+}$, $Dy^{3+}$ and $Ho^{3+}$) will not promote electron transfer via Ir-O-Ir and therefore the iridates remain insulating, and the larger cation promotes metallic nature [25]. The ionic radii of A cation also play a significant role in MIT: the MIT observed in $Nd_2Ir_2O_7$ (Nd3+: 1.12 Å) is at 36 K, for $Sm_2Ir_2O_7$ (Sm3+: 1.098 Å) and $Eu_2Ir_2O_7$ (Eu3+: 1.087 Å) it is 117 K and 120 K, [24, 26-27] respectively, which will be further discussed.

### b. B-site electronic configuration on the electrical property of pyrochlore

Table I shows some of the pyrochlores and their associated electrical property. It may be noted that, if B-site ion has unpaired electron then the compound exhibits fascinating electrical properties like Metal to Insulator Transition (MIT), Superconducting (SC), Magneto-resistive (MR) and Metal to semiconductor transitions (MST). The compound $Tl_2Mn_2O_7$ (3d pyrochlore) is exceptional as it shows MR property due to the hybridisation of *3d*-orbital of Mn with *6s* orbital of Tl [28-30]. It should also be noted that Tl-*6s* orbital and Mn-*3d* orbital energy bands are narrow which aids the MR property [30].

Pyrochlore oxides having *4d* or *5d* transition metal ions in the B-site usually consociate with MIT along with the magnetic properties of the A-site ion [24]. MIT in the pyrochlore family is relatively due to the electron-electron interaction of B-site ions that play an important role in electrical and magnetic transitions. As discussed earlier, $A_2Ir_2O_7$ (A = Pr, Nd, Sm and Eu) compounds were found to exhibit insulator behaviour at low temperatures and metal above MIT temperature, while $A'_2Ir_2O_7$ (A' = Gd, Tb, Dy) exhibit semiconducting at low temperature and metal above transition temperature [25]. This type of electrical property is due to the hybridisation of oxygen ions of the $t_{2g}$ energy level of $Ir^{4+}$ [31]. $Nd_2Ir_2O_7$, $Sm_2Ir_2O_7$, $Eu_2Ir_2O_7$ have sharp MIT at 36, 117 and 120 K, respectively. Similarly, $Cd_2Os_2O_7$, $Pr_2Mo_2O_7$ are also associated with the electrical transition in the low-temperature regime. Except for $Cd_2Os_2O_7$, MIT is peculiarly associated with magnetic ordering, a second-order phase transition confirmed from heat capacity measurements [16].

$Nd_2Ir_2O_7$ and $Sm_2Ir_2O_7$ also show MIT in the range of 50 to 150 K, which is due to the hybridisation of *5d* electron of $Ir^{4+}$. Even molybdenum pyrochlore ($A_2Mo_2O_7$) series also show a sharp transition at low temperatures, due to $Mo^{4+}$ interaction with oxygen and A-rare earth ion. As mentioned above, a new pyrochlore compound – $Hg_2Ru_2O_7$ also exhibits a MIT with magnetic transition [19, 32]. Studies on the single crystal of pyrochlore compounds further confirm MIT with electrical and magnetic transitions, as seen reported in $Sm_2Mo_2O_{7-\delta}$, $Gd_2Mo_2O_{7-\delta}$ and $Ho_2Mo_2O_{7-\delta}$ [33].

Among all the pyrochlore family's, $A_2Ti_2O_7$ is exceptional because the compound does not show any sign of electrical transition due to the vacant *3d* orbital of $Ti^{4+}$ ion. By comparing $Ti^{4+}$ ion with rest of the B-site electronic configuration as in $Ru^{4+}$, $Ir^{4+}$, $Os^{4+}$ and $Mo^{4+}$, $Ti^{4+}$ has vacant *d* ($3d^0$) configuration. Roth et al., in 1956 first reported the synthesis of lanthanide titanates like $Ln_2Ti_2O_7$ (Ln: Sm, Yb and Y) [34] and L.H. Brixner in 1964 reported the synthesis of $Tb_2Ti_2O_7$ and other pyrochlore oxides [35]. The electrical resistivity of most of the lanthanide titanates, $La_2Ti_2O_7$, $Nd_2Ti_2O_7$, $Sm_2Ti_2O_7$, $Ho_2Ti_2O_7$ and $Er_2Ti_2O_7$ were measured (by Brixner in 1964) upto 1000 °C which showed *p*-type semiconducting nature at the higher range of temperature. The resistivity of *3d* based pyrochlores (except $Tl_2Mn_2O_7$) at room temperature are in the range of insulators which is due to the absence of free electrons in B-cation [35]. However, the electrical resistivity of $Tb_2Ti_2O_7$ is still considered a fruit under the ice-berg, as interesting properties are further expected in these pyrochlore oxides.

$Tb_2Ti_2O_7$ is novel because of the (i) absence of long-range ordering even down to 17 mK [6, 36] (ii) absence of superconducting behaviour unlike $Cd_2Re_2O_7$ (iii) strong magnetic anisotropy along [111] axis [37] (iv) *co-operative Paramagnetism*, which is contrary to the *long-range spin order*, a property of spin ice and spin glass [6] (v) presence of glassy behaviour even at 200 mK [38] and (vi) absence of lattice deformation which makes $Tb_2Ti_2O_7$ a perfect pyrochlore [39]. Also, $Tb_2Ti_2O_7$ is not associated with any structural transition upto 600 K, suggesting that this compound is a perfect pyrochlore lattice. This is proved by S.W.Han.et.al through the neutron powder diffraction (NPD) studies [40].

The effect of *d*-electron (B site) in this compound is unexplored, due to the non-degenerate ground state of $Tb^{3+}$ similar to $Ti^{4+}$ in $Ho_2Ti_2O_7$ [41]. Pyrochlore iridates and molybdate like $R_2Ir_2O_7$ and $R_2Mo_2O_7$, where R =Y, Ho, Tb, Dy, Gd, Sm, Eu, Nd, Yb, show the lanthanide effect due to Ir electron density from *f* and *d* orbitals that play an important role in magnetic and electrical transitions [24, 42]. But the major contribution will be due to *f* electron. But in $Tb_2Ti_2O_7$, $Ti^{4+}$ has empty *d*-electron density. Therefore, the absence of *f*-orbital electron prompts the investigation of the property of vacant *d*-electron, in detail.

This article provides the first information on

- ➢ The formation of $Tb_2Ti_2O_7$ phase at a relatively low synthesis temperature, prepared by high energy ball milling and subsequent firing.
- ➢ Metal-like to insulator transition (MTI) in Ti based pyrochlore at 603 K supported by magnetic, heat capacity and differential scanning calorimetry measurements.
- ➢ A broad change in the magnetic moment from 696 K evidenced by heat capacity and differential scanning calorimetry.

II. **Experimental**

Preparation of phase pure pyrochlore oxides is difficult due to the formation of secondary oxygen deficient phases like $TbTi_2O_5$, and some oxygen rich phases like $Tb_4O_7$. First attempt to prepare $Tb_2Ti_2O_7$ was carried out by Brixner in 1964 using wet ball milling by mixing the corresponding oxides and subsequent firing at 1050 °C for 10-14 hour followed by firing again at 1350 °C for 10 to 14 hour [35]. Thus, the pyrochlore oxides are prepared usually by solid state reaction method by firing the corresponding oxides of rare earth and transition metals in the temperature range of 1350 °C (1623 K). For example, some oxides like $Ln_2Ir_2O_7$, Ln: Pr, Nd, Sm and Eu, were prepared in sealed tubes for several days (more than 5 days) with intermediate grinding, whereas $Tl_2Mn_2O_7$ was prepared using high pressure techniques [24, 34, 40, 43-46].

In the present study, polycrystalline $Tb_2Ti_2O_7$ is prepared by high energy ball milling (dry medium) followed by firing the powder in a furnace. $Tb_4O_7$ (Alfa Aesar, purity – 99.99 %) and $TiO_2$ (Merck, Purity – 99.9 %) taken in stoichiometry are milled in zirconia vial and balls at 500 rpm for 5 hour in a Fritsch Pulverisette - 7 planetary micro ball mill with a ball to powder ratio of 5:1. The sample is then scraped and subsequently fired at 950 °C (1223 K) in an alumina crucible for 5 hour. The equation representing the reaction is

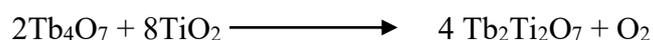

$$2Tb_4O_7 + 8TiO_2 \longrightarrow 4\ Tb_2Ti_2O_7 + O_2$$

The prepared sample is analysed using X-ray diffraction (XRD) study at room temperature (D8 Advance, Bruker) using Cu-$K_\alpha$ radiation of wavelength 1.5406 Å with 2θ ranging from 10° to 80° at a scan rate of 0.02° steps per second. The electronic state of Tb and Ti is determined through X-ray photoelectron spectroscopy using Al-$K_\alpha$ radiation (Omicron nanotechnology) and the low temperature and high temperature magnetisation measurements are performed in a Lakeshore 7410 model – Vibrating sample magnetometer at 1000 Oe and 7000 Oe, respectively. Solartron 1260 impedance/gain phase analyser is used to investigate the electrical property with Pt-probes and heat capacity measurement was performed in Neztech DSC-204F1 ($N_2$ atmosphere) (5 K/min). Differential Scanning Calorimetry (DSC) measurement is performed in DSC Polyma 214 Neztech under nitrogen atmosphere (10 K/min).

### III. Results and discussion
#### a. Structural analysis – X-Ray diffraction

The crystallographic structure is analysed by Rietveld refinement of the XRD pattern using the FULLPROF program. Figure 1 shows the diffraction peaks fitted using Pseudo-Voigt function and the background with 12- coefficient polynomial function. The compound adopts cubic structure with space group *Fd-3m* and the unit cell parameters and reliability factors obtained from the fit are tabulated in Table II. The refined unit cell parameters *are a = b = c = 10.1504(2) Å* with the cell volume of 1045 Å$^3$, in agreement with the reported values[7, 35, 40, 47-48]. The density of the prepared sample is calculated using lattice parameters from XRD data using the relation

$$\rho = \frac{\sum A}{N_A V}$$

Where ρ is the density of the compound (g/cm³), $\sum A$ is the product of the number of atoms in the unit cell to the atomic weight of the compound, $N_A$ – Avogadro number and V is the volume of the unit cell (1045 Å³ – calculated from XRD). By substituting the value, density is found to be 6.6 g/cm³ [47-48] which is in agreement with the values reported.

The bond lengths and bond angles are tabulated in Table III. Tb ions bond with O1 and O2 oxygen ions in the axial and equatorial position. The axial Tb-O2 (2.508(1) Å) bond length is found to be larger than that of the equatorial Tb-O1 (2.197(2)) Å. This is due to a larger spatial occupancy of Tb towards axial than the other. But in the case of Ti-O2, all the six oxygen atoms are equidistant with a bond length of 1.959(1) Å. This is in agreement with S.W. Han et.al., from the neutron powder diffraction studies, upto the second decimal [40].

### b. Electronic Analysis – X-ray Photoelectron spectroscopy

The XPS spectra are shown in figure 2. The survey spectrum in figure 2.a shows the binding energy peaks corresponding to $Tb^{3+}$, $Ti^{4+}$ and $O^{2-}$ ions. The peak obtained for Ti (fig 2.b) is deconvoluted and the corresponding binding energies obtained are 464 and 458.3 eV. It is well known that the peak at 464 and 458 eV correspond to $Ti^{4+}$ oxidation states $2P_{1/2}$ and $2P_{3/2}$ respectively[49-52]. Figure 2.c shows spectrum of Tb $3d_{3/2}$ and $3d_{5/2}$ corresponding to 1276 and 1241 eV respectively. High resolution XPS spectrum of Tb, in fig 2.d, shows the peak corresponding to 148.8 eV due to the $4d$ electron [53]. Therefore, comparison of the XPS binding energy with standard data, confirms Tb, Ti in 3+ and 4+ oxidation states, without any oxygen deficiency.

### c. Low Temperature Magnetic studies - from Vibrating Sample Magnetometer

The dc-magnetic susceptibility measurement is carried out at a constant magnetic field of 1000 Oe. Figure 3 shows the inverse susceptibility as a function of temperature. The data is fitted using Curie-Weiss law where the parameters were adopted from Gingras et al. ie., C = 23.0 emu K mol⁻¹, $\theta_{CW}$ = -18.9 K [2] and the effective Bohr magnetron ($P_{eff}$) is calculated using electronic configuration ($^7F_6$) of $Tb^{3+}$ ion as 9.7μ$_B$ per $Tb^{3+}$ ion and the plot shows a good agreement with experiment and Curie Weiss law. This confirms that the Curie Weiss law describes the experimental data till 50 K as reported by Gingras et al. and the negative Curie temperature indicates the interaction as anti-ferromagnetic in nature [2]. Therefore, dc susceptibility further reinforces the magnetic phase purity of the prepared $Tb_2Ti_2O_7$ sample.

### d. Transport Analysis – MIT in $Tb_2Ti_2O_7$

The prepared pyrochlore, $Tb_2Ti_2O_7$ is compressed into a pellet of thickness 0.9 mm and diameter of 8 mm and heat-treated at 500 °C for 5 hour to remove the moisture adsorbed. The pellet is subjected to impedance analysis in the frequency range of 10 MHz to 1 Hz. Dielectric constant, at room temperature, is observed as 40.8 which is very close to the reported values [35]. Resistivity (resistance) of the sample is obtained by fitting Z' vs Z" (real and imaginary part of impedance). Figure 4 shows the resistivity of $Tb_2Ti_2O_7$ (normalised to 603 K) as a function of temperature and the inset gives resistivity vs temperature. The plot reveals the following

I. The resistivity of the sample increases from 403 K to 603 K (i.e.) $\rho \alpha T$, and shows positive temperature coefficient of resistivity ($d\rho/dT > 0$) which confirms the metal-like behaviour of the sample.
II. Beyond 603 K, the resistivity of the sample decreases (i.e.) $\rho \alpha 1/T$, and $Tb_2Ti_2O_7$ shows a negative temperature coefficient of resistivity which is a behaviour typical of insulators.
III. The change in the order of resistivity of the transition is approximately $10^2$ which is also consistent with Nd, Sm and Eu iridate pyrochlores.

In general, the temperature at which the resistivity reaches the maximum and then declines with an increase in temperature is defined as the metal to insulator transition temperature ($T_{MIT}$). In our case, the trend of resistivity vs temperature and the temperature coefficient of resistivity shows a behaviour typical as that of metal-like to insulator transition ($T_{MTI}$).

Figure 5 (Z" spectra) shows the Z" (imaginary part of impedance) vs log $f$ where a suppression in the peak from 403 K to 603 K is seen confirming the metallic nature of the sample. Beyond MTI (603 K), the peak shifts towards the high frequency regime – which is characteristics of an insulating. The inset of fig 5 shows unity in $\tau$ (relaxation time) which decreases by 1/100$^{th}$ of order beyond MTI. Figure 6 shows the electric modulus spectra from which it is inferred,

I. a decrease in magnitude in M" vs log $f$ till 603 K and
II. a shift in peak towards a higher range of frequency from 696 K

Both the modulus (electric and impedance) spectra show a decrease in magnitude till MTI (603 K) and a shift in peak towards higher frequency beyond MTI.

Similar kind of electrical behaviour (MIT) is also found in other pyrochlores $Nd_2Ir_2O_7$, $Sm_2Ir_2O_7$, $Eu_2Ir_2O_7$, $Gd_2Ir_2O_7$, $Tb_2Ir_2O_7$, $Dy_2Ir_2O_7$, $Ho_2Ir_2O_7$, $Nd_2Mo_2O_7$, $Hg_2Ru_2O_7$, $Tl_2Ru_2O_7$ and $Tl_2Mn_2O_7$ which is due to $d$-orbital electron in B-transition metal ions like $Ir^{4+}$, $Ru^{4+}$ and $Mn^{4+}$ respectively [18-19, 24, 27, 54-55]. MIT is one of the common properties exhibited by iridium, and ruthenium pyrochlore families, but no reports are available in Ti-based pyrochlores. But the major difference in $Tb_2Ti_2O_7$ is, the temperature of MTI is around 600 K. To the best of our knowledge this is the first report for electrical behaviour in a Ti-based compound of the pyrochlore family. Even though the pyrochlore family exhibits a series of MIT in the low temperature regime, no reports are available for the transition in the high temperature regime in spin glass, spin ice and spin liquid materials (pyrochlores).

The observed MTI in $Tb_2Ti_2O_7$ is due to hybridisation of the vacant $3d$ orbital of $Ti^{4+}$ ion with $O^{2-}$ ion. Due to hybridisation, there is a relative increase in the electron density of $Ti^{4+}$ ($3d$) similar to other pyrochlore compounds. As a result, the bandgap of the material decreases which in turn decreases the resistivity with the rise in temperature. On the other hand, B cation is in six-fold coordination with oxygen due to which the $d$-state of B cation (empty $d$-state of $Ti^{4+}$ ion) show a significant covalent interaction with $2p$ state ion. Further, $6p$ orbital is absent in $Tb^{3+}$ ion and hence hybridisation of $Ti^{4+}$ and $O^{2-}$ ions alone are possible. Therefore, the transfer of electrons from $2p$ orbital of O to vacant $3d$ orbital of $Ti^{4+}$ ion (figure 7), which has relatively closer energy levels, is responsible for the metallic nature [56] observed. Figure 8 shows the local oxygen environment of Tb, Ti and $Tb_2Ti_2O_7$. Tb and Ti have 8 and 6 oxygen ion co-ordination respectively. At higher temperatures, due to thermal vibration the bond distance between Tb and O1 increases. O1 of Tb repels O2 of Tb thereby decreasing the bond length between Ti and O2- facilitating the electron transport from O2 and $Ti^{4+}$ ions, which is responsible for the decrease in the resistivity of the compound. The above discussion is also consistent with the bond length and angle shown by S.W. Han et. al., using NPD taken at 45 K.

The bond length of Ti and O2 reported is 1.9699 Å at 45 K (Ref no: 38). But in the present work, at 300 K, it is found to be 1.9590 Å (Table I). Also, the bond angle increases from 132.12° (ref 38) to 132.70° which favours the hybridisation of $Ti^{4+}$ ion with $O^{2-}$ ion. Also, the bond length of Tb and O1 increases from 2.1939 Å (ref 38) to 2.1978 Å which supports our claim on MTI. We strongly believe that Ti-O2 bond length further decreases above 600 K that favours the observed MTI. The MTI in $Tb_2Ti_2O_7$ is the only observation so far in the high temperature regime, whereas all other compounds exhibit electrical transition in the low temperature regime as discussed in the previous sections.

The stability of this compound is already verified by Han et.al using neutron powder diffraction and X-ray absorption fine structure and confirm that $Tb_2Ti_2O_7$ is not associated with any structural transition in the range of 4.5 to 600 K. More than these, to confirm that the observed MTI is not due to oxygen vacancy at elevated temperatures and also to confirm that the compound is not associated with any secondary phase, the prepared $Tb_2Ti_2O_7$ is subjected to Thermogravimetric analysis (TGA) (inset of fig 6) in the temperature range of 450 K to 800 K. Two affirmative results: (i) no weight loss (from TGA) in the entire range of measurement confirms the absence of any oxygen vacancy (ii) no additional peaks in XRD pattern, other than the characteristics, are observed showing the absence of secondary phase in the prepared compound.

e. **High Temperature – Magnetic studies**

The prepared $Tb_2Ti_2O_7$ is subjected to high temperature magnetic measurement from 400 to 900 K at a constant magnetic field of 7000 Oe (Figure 9). Two major conclusions are drawn: (i) At 600 K there is a small difference in the magnetisation, which is clearly shown in the inset as the first derivative of the magnetisation – the change in the magnetic moment is associated with MIT and (ii) above 696 K, the peak is broad, diffused and consistent with the peak shift obtained in M" vs $\log f$ plot as shown in figure 6. This broad change in magnetisation ($T_M$) from 696 K may be due to the induced magnetic moment of $Tb^{3+}$ that needs to be further analysed. Therefore, the high temperature magnetic measurement confirms (i) change in the magnetic moment is associated with MTI (603 K) and (ii) the subsequent broad change in magnetisation from 696 K may be due to $Tb^{3+}$ rare earth ion. However, in addition to hybridisation of Ti-O, any possible interactions on electrical transport and change in magnetisation need to be understood by the theoretical investigation.

f. **Heat Capacity and Differential Scanning Calorimetry**

Specific heat capacity measurements in the range from 450 to 800 K is shown in figure 10. A broad hump at 603 K supports the MIT observed in electrical property and another peak at 696 K supports the broad change in magnetisation observed in magnetic measurements. Differential Scanning Calorimetry (DSC) is also performed (a sensitive tool to detect transitions) and is shown in figure 11 that confirms MIT through the sharp peak at 603 K, and another sharp peak at 696 K indicative of the observed change in magnetisation. Thus, the Heat capacity and DSC confirm both the (i) MIT at 603 K and (ii) the broad change in magnetisation from 696 K.

## IV.  Conclusion

To conclude, a cubic pyrochlore $Tb_2Ti_2O_7$ is synthesised by solid state reaction method at a relatively low phase formation temperature effected by the prior high energy ball milling of the precursors. Metal-like to insulator transition (MTI) (through electrical resistivity) is found at 603 K. Relaxation time and electric modulus plots (real and imaginary) confirm MTI. DC magnetic susceptibility, heat capacity and differential scanning calorimetry measurements prove the MTI observed. Absence of thermal hysteresis indicates the second order phase transition. The MIT in $Tb_2Ti_2O_7$ is a consequence of *d*-orbital, unlike other pyrochlore oxides. This is the first report on Ti-based pyrochlore oxides which shows MTI driven by the empty 3d orbital of $Ti^{4+}$ and a subsequent broad change in magnetisation from 696 K. This study opens up new avenues for further research in Ti-based pyrochlore oxides in the electrical and magnetic pathway.

**Acknowledgement**

BSK thanks DST-INSPIRE for the award of SRF (IF-140582) and Mr. B. Soundararajan for his support in the measurements. NCNSNT, University of Madras, is acknowledged for XPS measurement. The University of Madras – Central Instrumentation Facility (GNR) is acknowledged for DSC and TGA measurements.

**Data availability statement**

The data that support the findings of this study are available from the corresponding author upon reasonable request.

Table 1: Materials belonging to the pyrochlore family exhibiting physical properties including superconductivity (SC), metal to insulator (MIT), magneto-resistance (MR) and semiconductor to metal (SMT) transition with the corresponding B metal electronic configuration and temperature (K).

| Compound | Electrons of B site ions | Property | T(K) | Ref |
|---|---|---|---|---|
| $Cd_2Re_2O_7$ | $5d^2$ | SC | 1 K | 14, 16, 28, 57 |
| $Nd_2Ir_2O_7$ | $5d^5$ | MIT | 36 K | 24 |
| $Sm_2Ir_2O_7$ | $5d^5$ | MIT | 117 K | 24 |
| $Eu_2Ir_2O_7$ | $5d^5$ | MIT | 120 K | 24 |
| $Gd_2Ir_2O_7$ | $5d^5$ | MIT | 127 K | 27 |
| $Tb_2Ir_2O_7$ | $5d^5$ | MIT | 132 K | 27 |
| $Dy_2Ir_2O_7$ | $5d^5$ | MIT | 134 K | 27 |
| $Ho_2Ir_2O_7$ | $5d^5$ | MIT | 141 K | 27 |
| $Nd_2Mo_2O_7$ | $4d^2$ | MR | 90 K | 58 |
| $Hg_2Ru_2O_7$ | $4d^3$ | MIT | 108 K | 19, 32 |
| $Tl_2Ru_2O_7$ | $4d^4$ | MIT | 120 K | 54 |
| $Tl_2Mn_2O_7$ | $3d^3$ | MR | 142 K | 55 |
| $Cd_2Os_2O_7$ | $5d^3$ | SMT | 225 K | 59 |

Table 2: Structural parameters obtained from Rietveld refinement of the XRD pattern of $Tb_2Ti_2O_7$. $\chi^2$ = 2.68, $R_p$ = 3.32, $R_{wp}$=2.09, $R_{exp}$ = 1.27, $a = b = c$ = 10.1504(2) Å, Cubic structure of space group *Fd-3m*.

| Label | Atom | Position | x | y | z | Occupancy |
|---|---|---|---|---|---|---|
| Tb1 | Tb | *16d* | ½ | ½ | ½ | 1 |
| Ti1 | Ti | *16c* | 0 | 0 | 0 | 1 |
| O2 | O | *48f* | 0.3274 | 1/8 | 1/8 | 1 |
| O1 | O | *8b* | 3/8 | 3/8 | 3/8 | 1 |

Table 3: Important bond length and bond angle obtained from Rietveld refinement of the XRD pattern of $Tb_2Ti_2O_7$, (at room temperature).

| Bond length (Å) | | Bond angle (°) | |
|---|---|---|---|
| Tb1 – O1 | 2.197 (2) | Tb1 – O1 – Tb1 | 109.38(3) |
| Tb1 – O2 | 2.508 (1) | O2 – Ti1 – O2 | 132.70(2) |
| Ti1 – O2 | 1.959 (1) | O2 – Ti1 – Tb1 | 90.00(1) |
| Ti1 – O1 | 3.777 (2) | Tb1 – O2 – Ti1 | 108.96(3) |
| Ti1 – Ti1 | 3.589 (2) | O1 – Tb1 – O1 | 180.00(1) |
| O2 – O1 | 3.621 (2) | | |

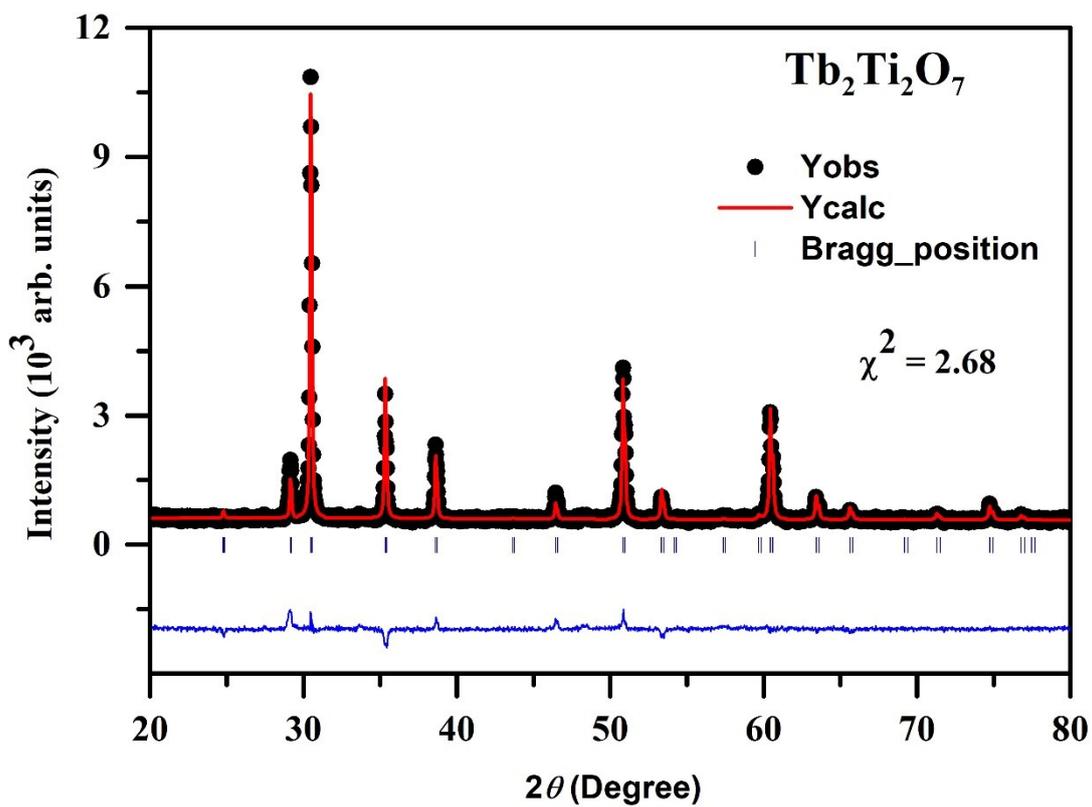

*Figure 1: Rietveld refinement of the XRD pattern of polycrystalline $Tb_2Ti_2O_7$ sintered at 950° C for 5 hour.*

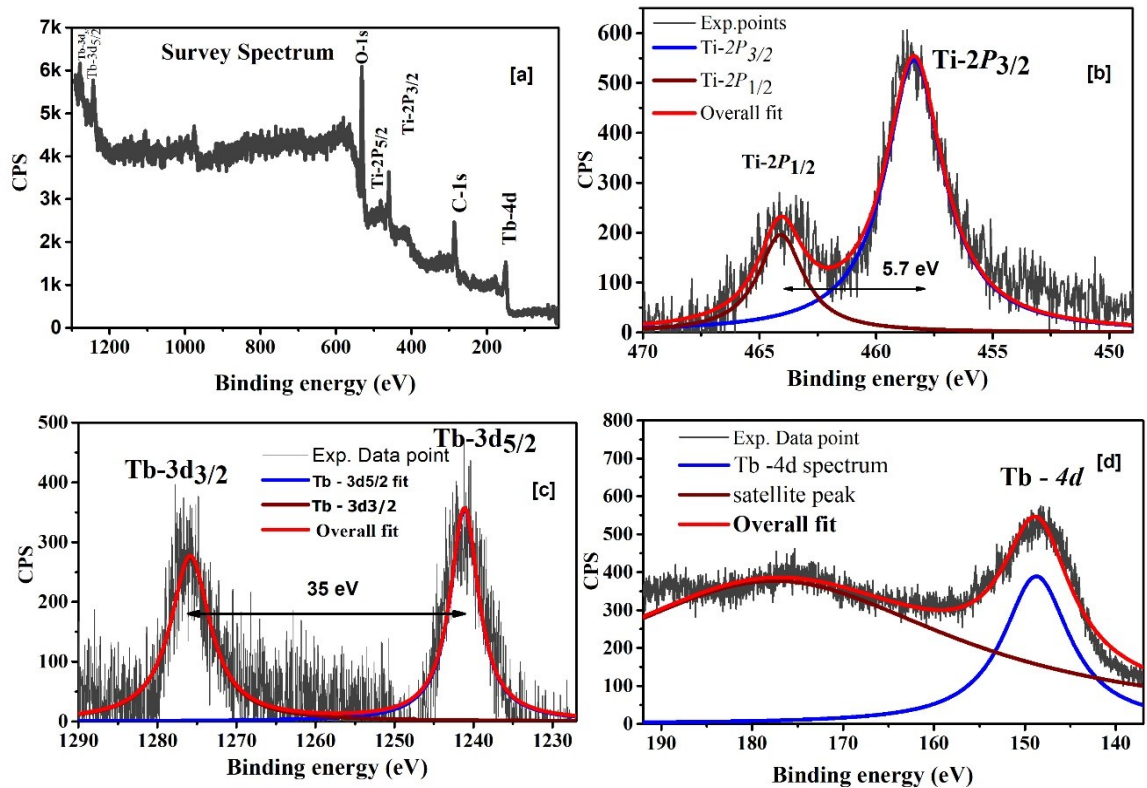

Figure 2: XPS spectra showing (a) survey spectrum of $Tb_2Ti_2O_7$ (b) binding energy spectrum of $Ti2p_{3/2}$ revealing +4 oxidation state of Ti (c) & (d) binding energy spectra of Tb- $3d_{5/2}$ and 4d confirming the +3 oxidation state of Tb

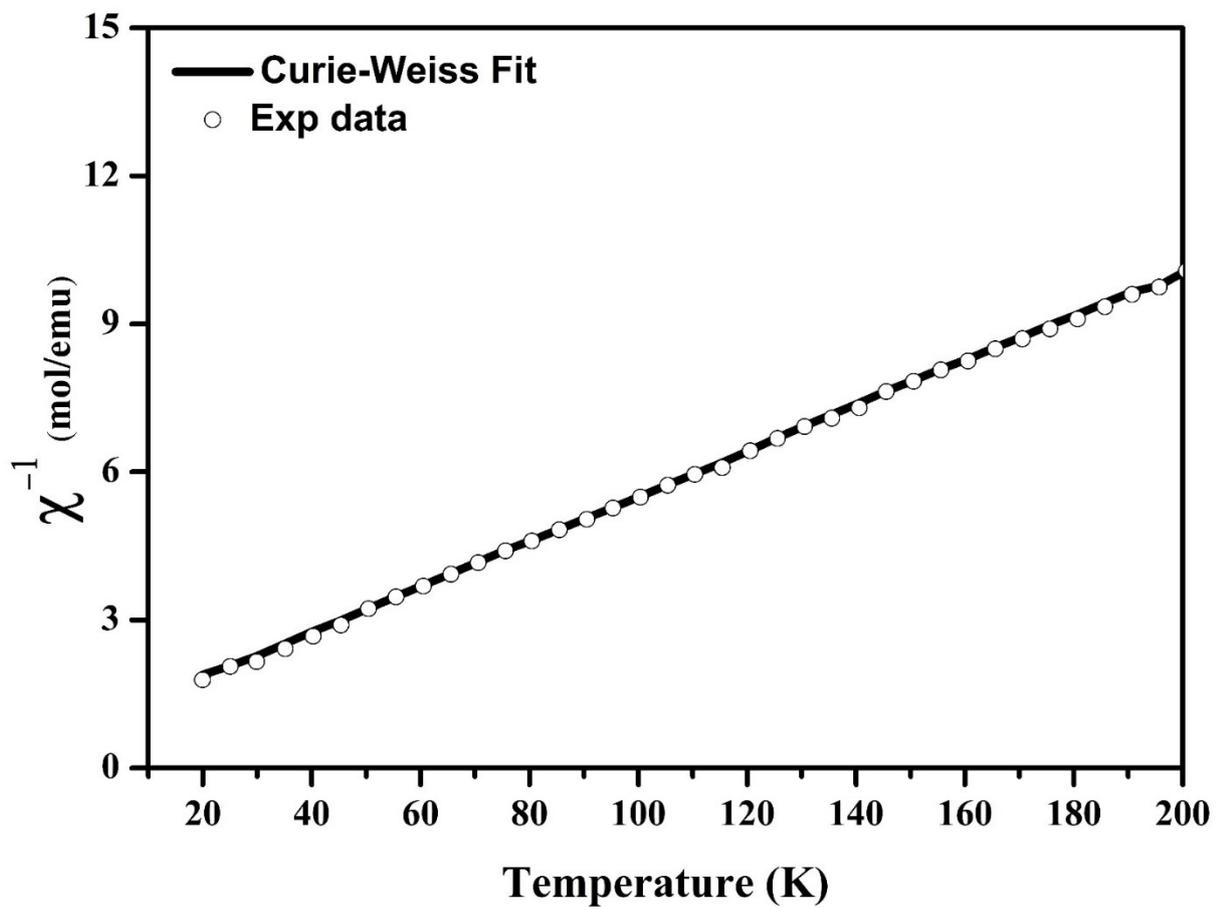

Figure 3: Temperature dependent DC susceptibility for Tb$_2$Ti$_2$O$_7$ at a constant magnetic field of 1000 Oe.

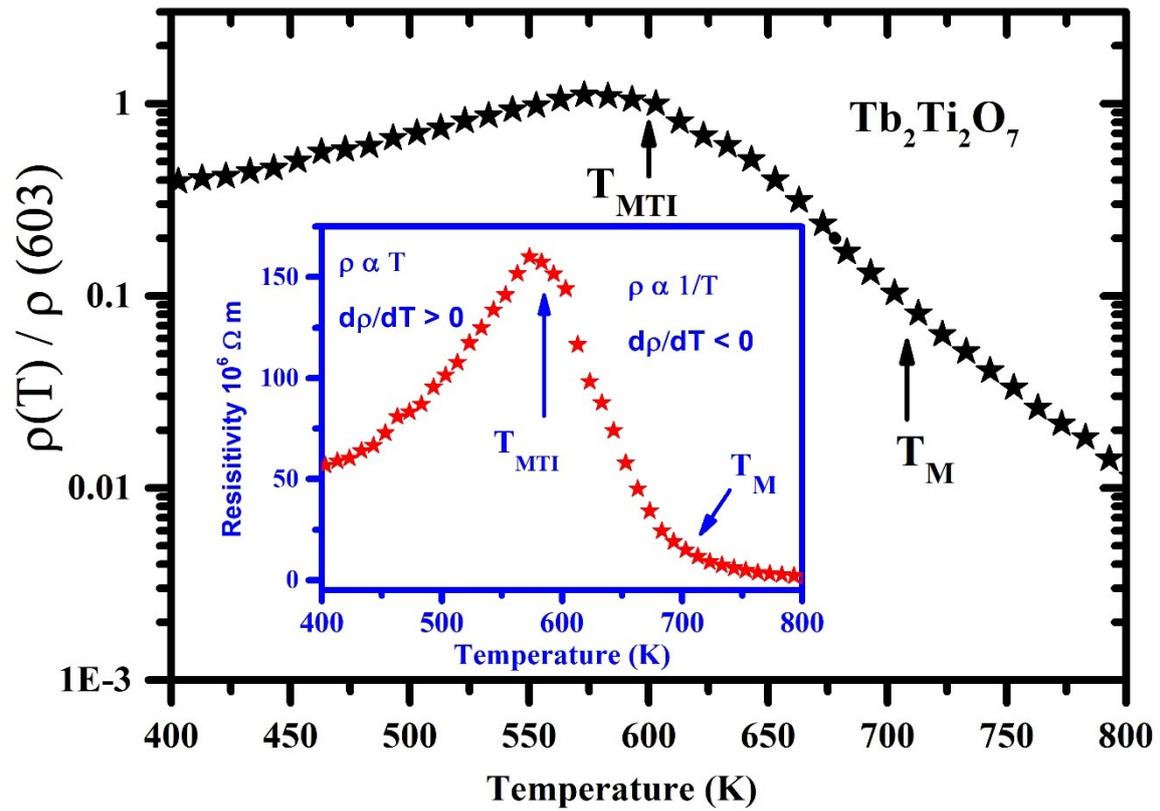

Figure 4: Variation of resistivity with temperature from 403 K to 803 K (normalised to 603 K) -$Tb_2Ti_2O_7$ which confirms metal-like to insulator transition (MTI). The inset shows the actual value of resistivity, the MTI at 603 K and $T_M$ indicates the broad change in magnetisation from 696 K.

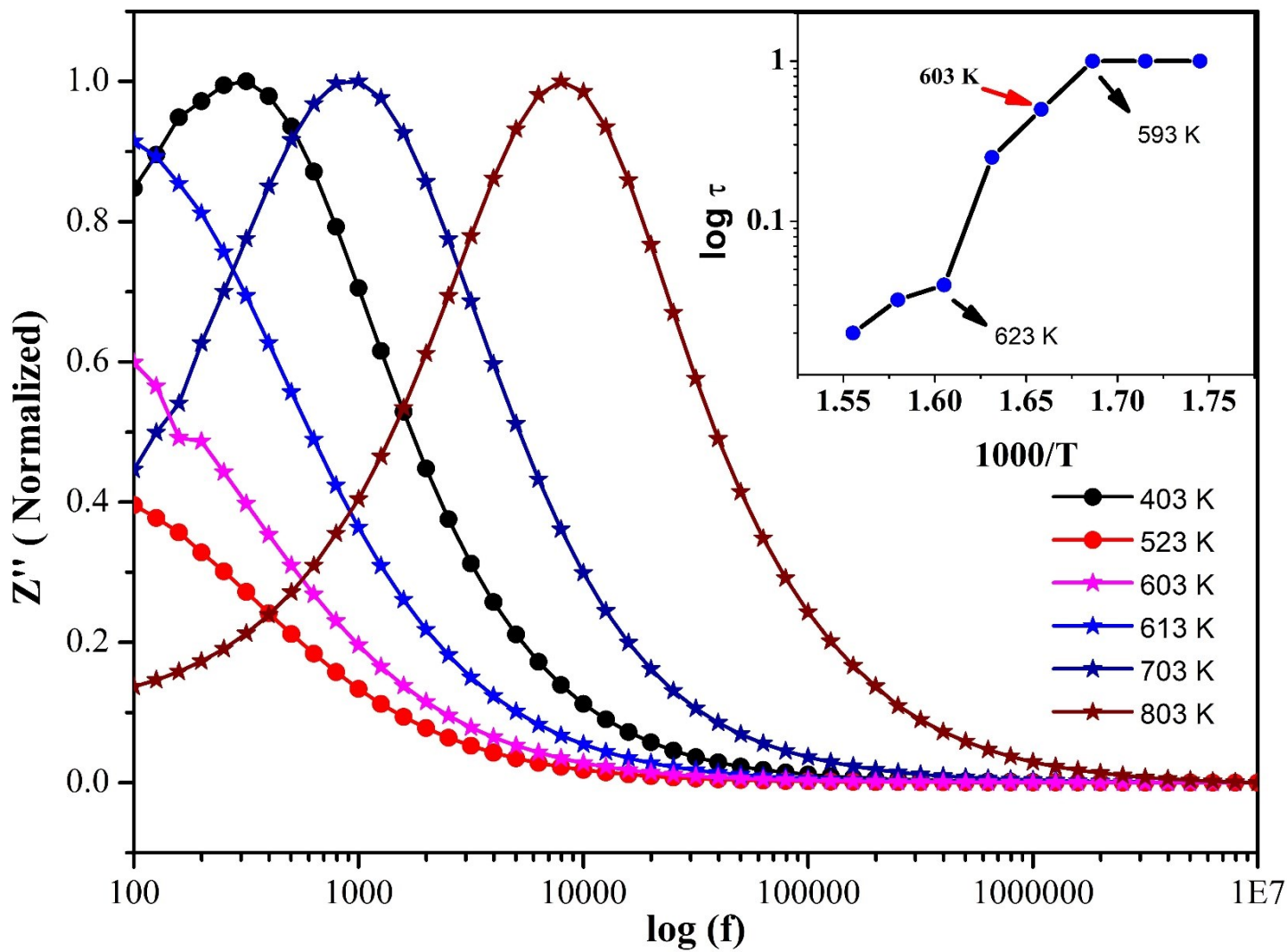

Figure 5: Imaginary spectrum (Z″) as a function of applied frequency - $Tb_2Ti_2O_7$. Inset shows the difference in relaxation time before and after the transition.

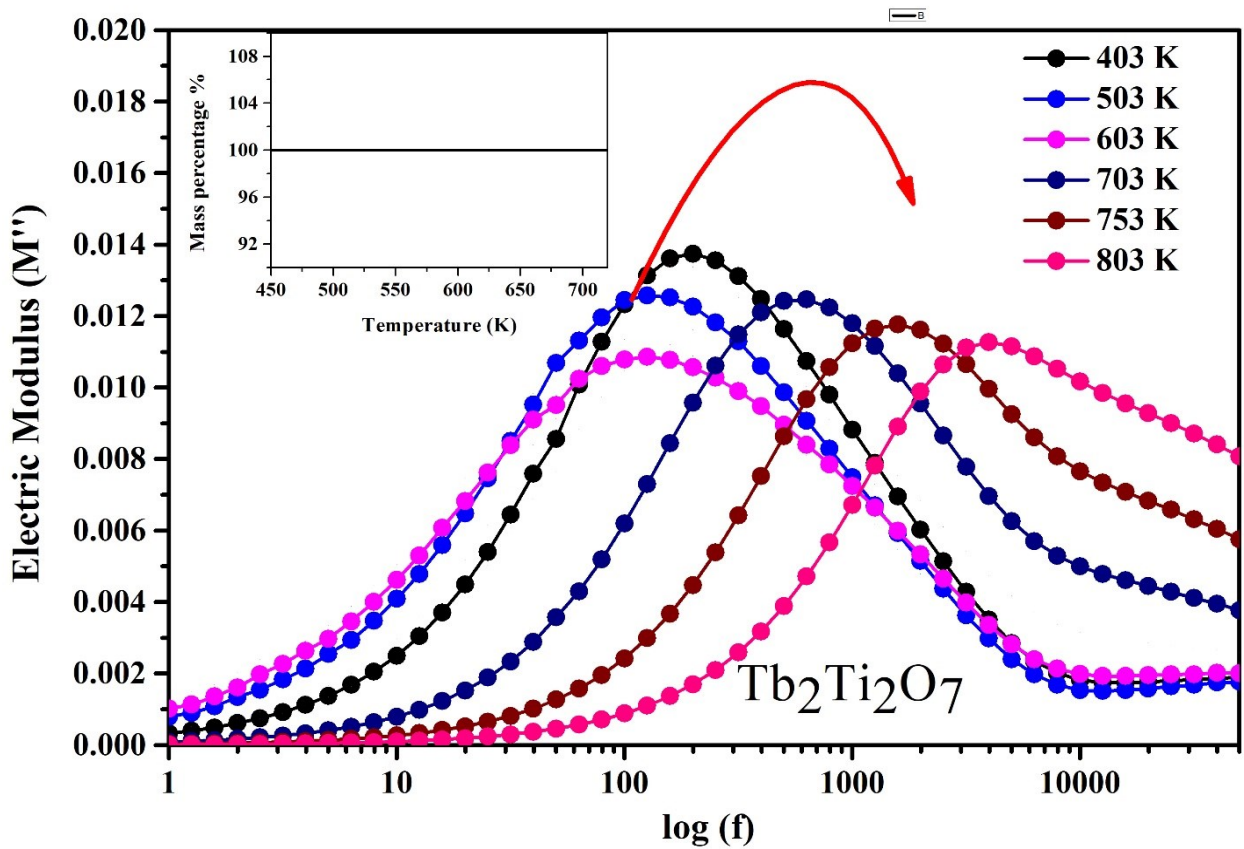

Figure 6: Electric modulus spectrum vs log (f) of $Tb_2Ti_2O_7$ showing the shift in peak and the corresponding region before and after MTI. (The red arrow indicates the shift in peak towards high frequency regime beyond 703 K). The inset shows the TGA plot of $Tb_2Ti_2O_7$ sample.

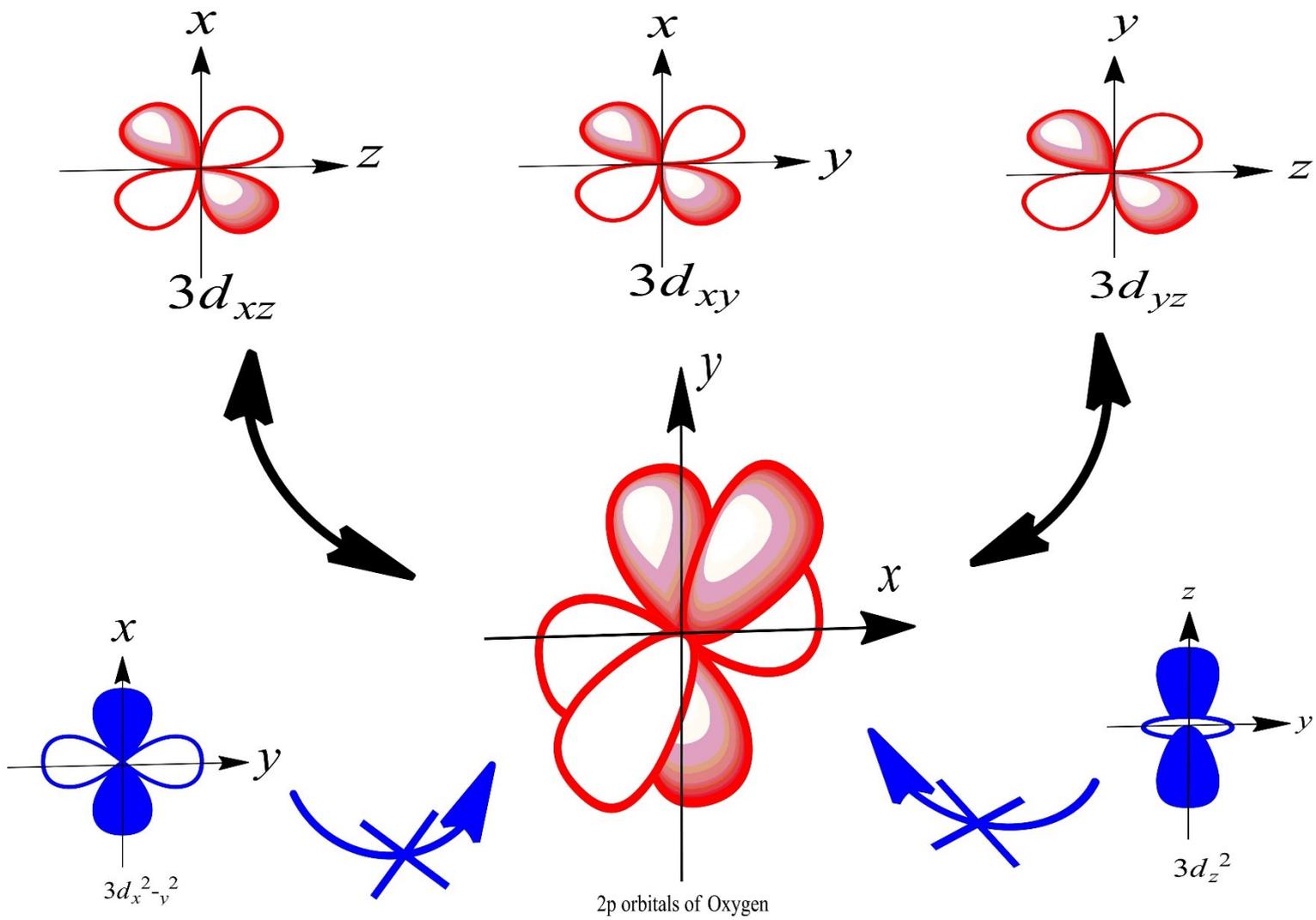

Figure 7: Schematic representation of the 3d orbital interaction of $Ti^{4+}$ with 2p orbital of $O^{2-}$ in $Tb_2Ti_2O_7$.

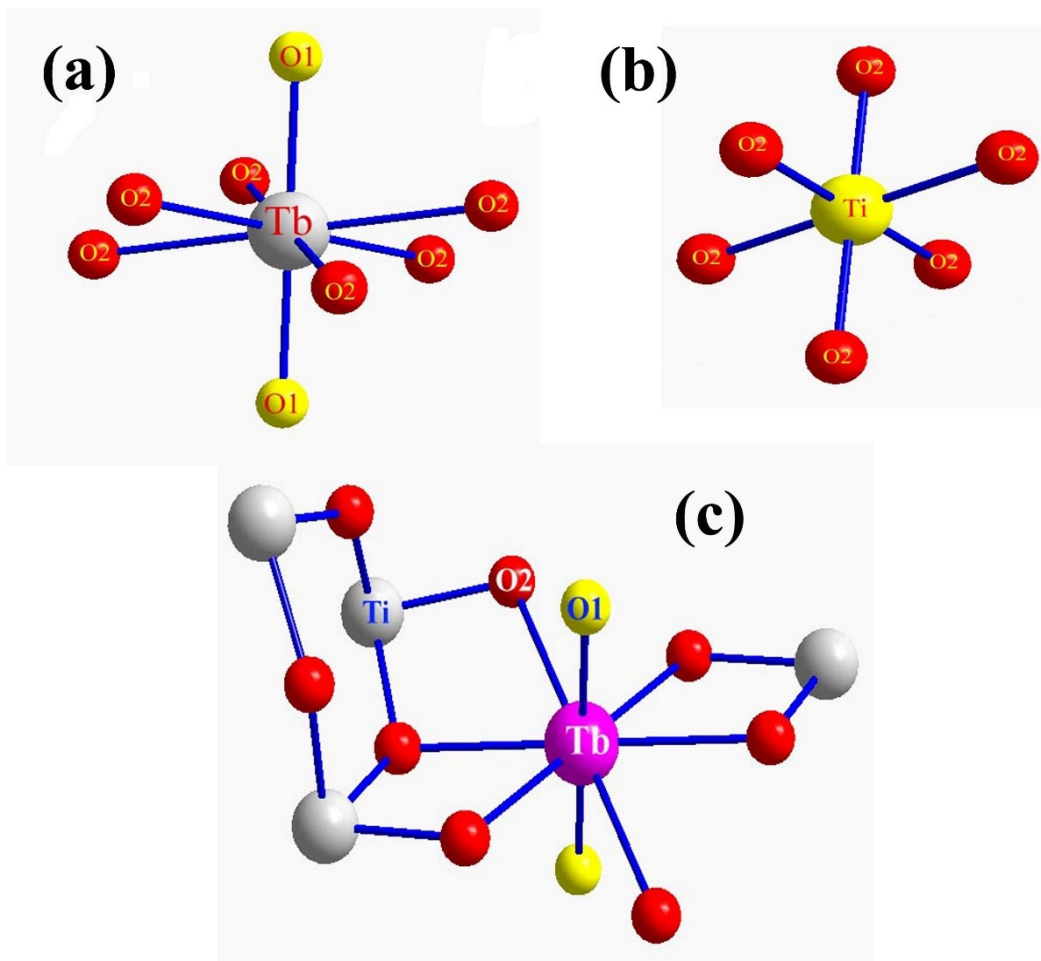

Figure 8: (a) Local environment of Oxygen ion around Tb, in $Tb_2Ti_2O_7$. O1 and O2 indicate the oxygen atom at axial and equatorial directions respectively, (b) Ti environment of Oxygen confirms that all the six oxygen ion are equidistant from Ti ion and (c) shows the overall picture of Tb and Ti with oxygen.

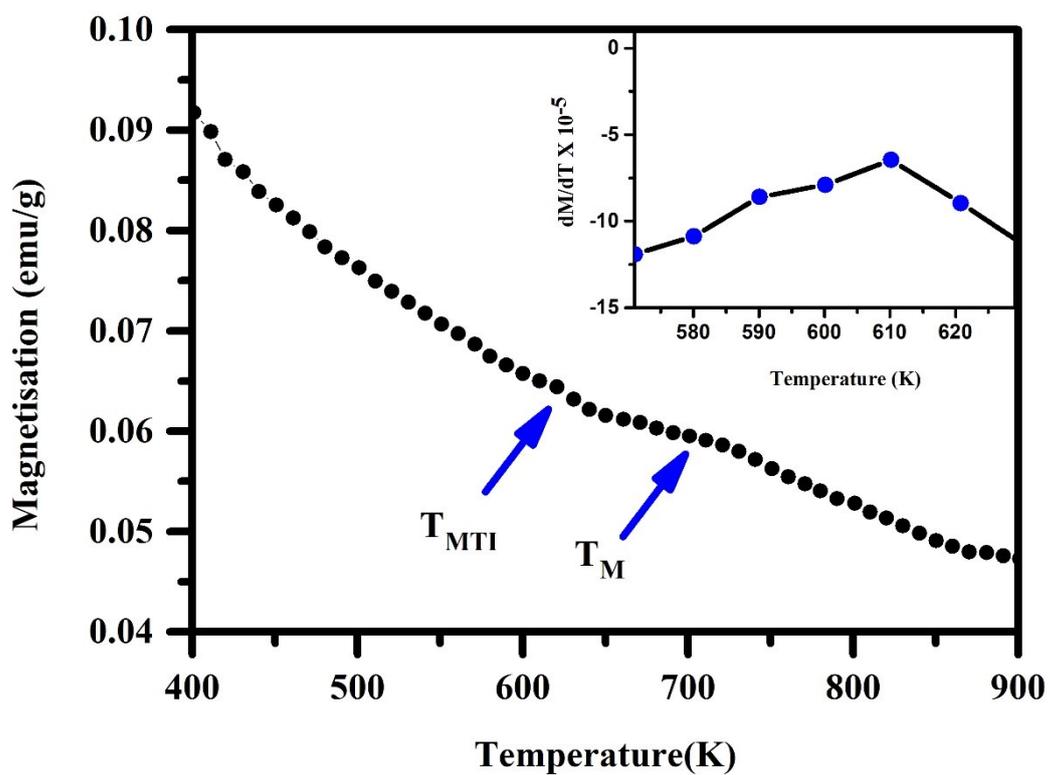

Figure 9: Temperature dependent magnetisation of $Tb_2Ti_2O_7$. Inset shows the first derivative of magnetisation with notable change at 600 K ($T_{MTI}$) and $T_M$ indicate the broad change in magnetisation from 696 K.

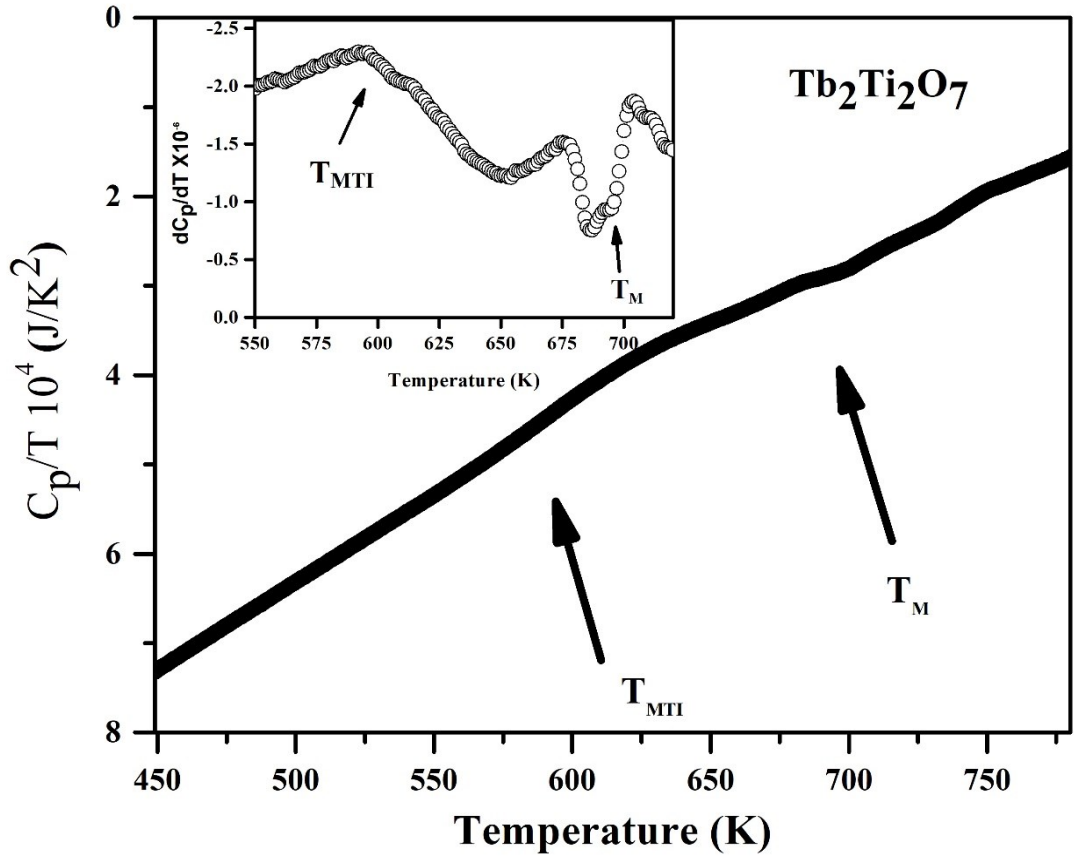

Figure 10: Heat capacity of $Tb_2Ti_2O_7$ from 450 to 800 K. $T_{MTI}$ indicates Metal-like to insulator transition around 600 K and $T_M$ indicates the broad change in magnetisation around 696 K. Inset shows the first derivative of heat capacity, which clearly shows a change in magnitude at 603 K (MIT) and 696 K.

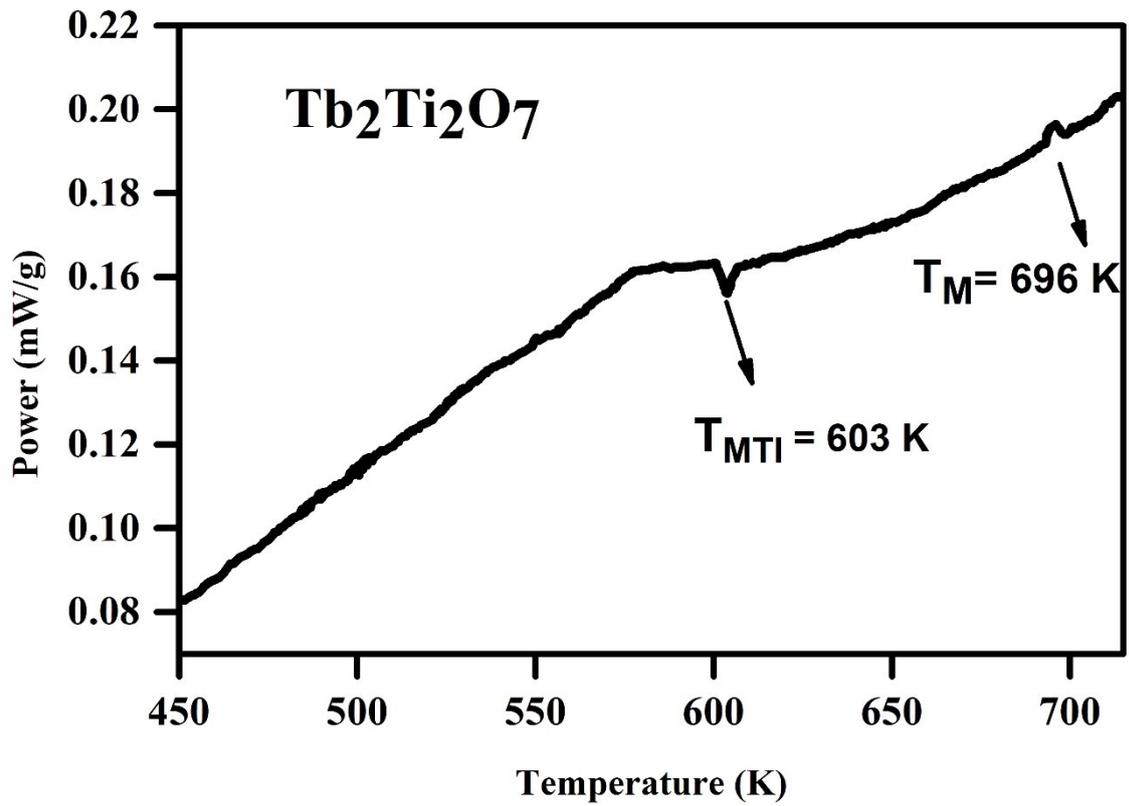

Figure 11: Differential scanning calorimetry measurement of $Tb_2Ti_2O_7$ from 450 to 725 K, which clearly shows metal-like to insulator transition (MIT at 603 K) and the broad change in magnetisation ($T_M$ at 696 K).